\documentclass[preprint,12pt,showpacs]{revtex4}

\usepackage[brazil, english]{babel}
\usepackage[latin1]{inputenc}
\usepackage{graphicx}
\usepackage{epsfig}
\usepackage{amssymb}
\usepackage{amsmath}
\usepackage[plainpages=false,pagebackref=false,pdftex]{hyperref}
\usepackage{ulem}

\begin{document}

\title{Twist Two Operator Approach for Even Spin Glueball Masses and Pomeron Regge Trajectory from the Hardwall Model}
\author{Diego M. Rodrigues$ ^{1} $}
\email[Eletronic address: ]{diegomr@if.ufrj.br}
\author{Eduardo Folco Capossoli$^{1,2,}$}
\email[Eletronic address: ]{educapossoli@if.ufrj.br}
\author{Henrique Boschi-Filho$^{1,}$}
\email[Eletronic address: ]{boschi@if.ufrj.br}  
\affiliation{$^1$Instituto de Física, Universidade Federal do Rio de Janeiro, 21.941-972 - Rio de Janeiro-RJ - Brazil \\
 $^2$Departamento de Física, Colégio Pedro II, 20.921-903 - Rio de Janeiro-RJ - Brazil}

\begin{abstract}
We compute the masses of even spin glueball states $ J^{PC} $, with $ P=C=+1 $, using a twist two operator from an AdS/QCD model known as the hardwall model, using Dirichlet and Neumann boundary conditions. Within this approach, we found that the glueball masses are comparable with those in literature. 
From these masses, we obtained the Pomeron Regge trajectories for both boundary conditions in agreement with experimental data available and other holographic models. 
\end{abstract}


\maketitle


\section{Introduction}
The AdS/CFT correspondence \cite{Gubser:1998bc,Maldacena:1997re,Witten:1998qj,Aharony:1999ti} relates a string theory defined on $ AdS_{5}\times S^5 $ space with a Yang-Mills theory with gauge group $ SU(N) $ with large $N$ in four dimensional Minkowski space. This Yang-Mills theory is conformal and supersymmetric with extended supersymmetry ${\cal N}=4$. 
At low energies, the string theory is represented by a supergravity theory in the $ AdS_{5}\times S^5 $ space-time, so this correspondence is also known as a gauge/gravity duality. 
Furthermore, this duality is such that when the string/supergravity theory is weakly coupled the super Yang Mills theory is strongly coupled. So, we can use this duality to investigate strongly coupled gauge theories in terms of a weakly coupled gravity theory.
These dualities can be interpreted as realizations of the holographic principle, since a theory in higher dimensions with gravity is described as another theory in lower dimensions without gravity \cite{'tHooft:1993gx, Susskind:1998dq}.

On the other side, $SU(3)$ Yang-Mills gauge theory coupled to quark fields known as QCD is the theory accepted to describe strong interactions. The quark and gluon interactions at high energies are perturbative but at low energies they fall in the nonperturbative regime. This regime is very important because it is related to difficult problems like quark and gluon confinement, hadronization, phase transitions, etc. The standard approach to QCD in this regime is to use some nonperturbative techniques as lattice field discretization, sum rules, etc. The strong interactions at low energies are also described by different phenomenological models. 

The advent of AdS/CFT correspondence or gauge/gravity dualities motivated the proposition of holographic models for QCD, known as AdS/QCD models. In these models some properties of hadronic physics are reproduced. In particular, the conformal symmetry should be broken in order to account for instance for hadronic masses. In this regard, the hardwall model was proposed to describe glueballs in a five dimensional slice of AdS space with a hard cutoff \cite{Polchinski:2001tt, BoschiFilho:2002ta, Polchinski:2002jw, BoschiFilho:2002vd}. Soon after, it was extended to mesons and baryons \cite{deTeramond:2005su}. As an alternative to break conformal invariance the softwall model was proposed introducing a dilaton field which behaves as a soft cutoff \cite{Karch:2006pv}.

In this work, we are going to study the properties of even spin glueball masses in the hardwall model and relate these masses with the corresponding Regge trajectory associated with the pomeron. A previous study was done \cite{BoschiFilho:2005yh} considering twist four operators. 
Here, we are going to use twist two operators starting with the action of a massive symmetric tensor representing a spin 2 field. From this action we obtain the equations of motion in terms of a scalar field. Then we extend this formulation for higher spins. The importance of the twist two operator in our formulation is that this twist is usually related to the pomeron in accordance with \cite{Brower:2006ea}.

This work is organized as follows: in Section II the Hardwall model is introduced, the twist two operator is constructed,  and we give a prescription to calculate the masses of glueball states with Dirichlet and Neumann boundary conditions. In Section III, we present a brief discussion of the Regge theory, our results for the glueball masses  and the Pomeron Regge trajectories for both Dirichlet and Neumann cases. 
Finally, our conclusions are presented in Section IV.

\section{The Hardwall model, twist two operator and glueball masses}

The starting point in our calculation is the massive symmetric second-rank tensor field action which will be related to the glueball state $2^{++}$. The $ D $-dimensional action for a massive spin-2 field in a curved spacetime consistent with the flat space limit is given by \cite{Buchbinder:1999be,Polishchuk:1999nh}: 
\begin{eqnarray} \label{TensorialGeneral_Action}
S[h_{\mu\nu}] = \dfrac{1}{2\kappa^{2}}\int d^{D}x\sqrt{|g|}\Big(\dfrac{1}{4}\nabla_{\mu}h\nabla^{\mu}h- \dfrac{1}{4}\nabla_{\mu}h_{\nu\rho}\nabla^{\mu}h^{\nu\rho} - \dfrac{1}{2}\nabla^{\mu}h_{\mu\nu}\nabla^{\nu}h \nonumber\\
+ \dfrac{1}{2}\nabla_{\mu}h_{\nu\rho}\nabla^{\rho}h^{\nu\mu} + \frac{\xi}{2D}\mathcal{R}h_{\mu\nu}h^{\mu\nu} + \frac{1-2\xi}{4D}\mathcal{R}h^2 - \dfrac{M^2}{4}(h_{\mu\nu}h^{\mu\nu} - h^2)\Big),
\end{eqnarray}
where $ h = g^{\mu\nu}h_{\mu\nu} $ is the trace, $ \mathcal{R} $ is the Ricci scalar and $ \xi $ is the only dimensionless coupling responsible for the nonminimality of interaction with the curved background. For AdS$_{D} $, the Ricci scalar is given by
\begin{equation} \label{Ricci_Scalar_AdS}
\mathcal{R} = -\dfrac{D(D-1)}{R^2}.
\end{equation}
The action \eqref{TensorialGeneral_Action} is the most general form of the action for a massive spin 2 field in curved spacetime of arbitrary dimension leading to consistent equations of motion plus constraints, which in turn are given by:
\begin{eqnarray} 
\nabla^{2}h_{\mu\nu} + 2\,\mathcal{R}^{\alpha\beta}_{\mu\nu}h_{\alpha\beta} - M²h_{\mu\nu} = 0 \label{E-Motion_TensorialGeneral_Action1}, \\
h^{\mu}_{\, \mu} = 0 \label{E-Motion_TensorialGeneral_Action2}, \\
\nabla^{\mu}h_{\mu\nu} = 0, \label{E-Motion_TensorialGeneral_Action3}
\end{eqnarray}
where $ \mathcal{R}^{\alpha\beta}_{\mu\nu} $ is the Riemann tensor. Eqs. \eqref{E-Motion_TensorialGeneral_Action1}, \eqref{E-Motion_TensorialGeneral_Action2} and \eqref{E-Motion_TensorialGeneral_Action3} took these simple forms because we set $ \xi = 1 $ in \eqref{TensorialGeneral_Action}. The theory is consistent for any value of $ \xi $, the only change is that the explicit form of constraints and dynamical equations of motion are more complicated than that of \eqref{E-Motion_TensorialGeneral_Action1}, \eqref{E-Motion_TensorialGeneral_Action2} and \eqref{E-Motion_TensorialGeneral_Action3}. Moreover, it is also possible to generalize the action \eqref{TensorialGeneral_Action} for the case of spin 2 massive field interacting not only with the gravity background but also with the scalar dilaton field. This field arises naturally in string theory as one of its massless excitations \cite{Buchbinder:1999be,Polishchuk:1999nh}. 

In our particular case, we are interested in the 5-dimensional version of \eqref{TensorialGeneral_Action} for AdS. Setting $ D=5 $, $ \xi = 1 $ and using \eqref{Ricci_Scalar_AdS} we have
\begin{eqnarray} \label{Tensorial_Action}
S[h_{\mu\nu}] = \dfrac{1}{2\kappa^{2}}\int_{AdS_5} d^{5}x\sqrt{|g|}\Big(\dfrac{1}{4}\nabla_{\mu}h\nabla^{\mu}h- \dfrac{1}{4}\nabla_{\mu}h_{\nu\rho}\nabla^{\mu}h^{\nu\rho} - \dfrac{1}{2}\nabla^{\mu}h_{\mu\nu}\nabla^{\nu}h \nonumber\\
+ \dfrac{1}{2}\nabla_{\mu}h_{\nu\rho}\nabla^{\rho}h^{\nu\mu} - \frac{2}{R^2}h_{\mu\nu}h^{\mu\nu} + \frac{1}{R^2}h^2 - \dfrac{1}{4}M_5^2(h_{\mu\nu}h^{\mu\nu} - h^2)\Big),
\end{eqnarray}
where $ M_5 $ is the five-dimensional mass.

The holographic model we adopted in this work is the hardwall model \cite{Polchinski:2001tt,Polchinski:2002jw, BoschiFilho:2002ta,BoschiFilho:2002vd,BoschiFilho:2005yh,Capossoli:2013kb}, in which a hard cutoff is introduced in the AdS space in order to break conformal invariance. The metric of AdS$ _5 $ space can be written in the form
\begin{equation}
ds^2 = \dfrac{R^2}{z^2}(dz² + \eta_{\mu\nu}dx^{\mu}dx^{\nu})\,,
\end{equation}
where $R$ is a constant identified with the AdS radius, $\eta_{\mu\nu}$ has signature $(-, +,+,+)$ and we are disregarding the $ S^5 $ part of the metric. The introduction of a cutoff in AdS space in this model implies that 
\begin{equation}
0\leqslant z\leqslant z_{max},
\end{equation}
where $ z_{max} $ can be related to the QCD mass scale
\begin{equation}
z_{max} \sim \dfrac{1}{\Lambda_{QCD}}
\end{equation}
and some boundary conditions have to be imposed in $ z=z_{{max}} $. In this work, we have used Dirichlet and Neumann boundary conditions.

Now, using the polarization given by eqs. \eqref{E-Motion_TensorialGeneral_Action2} and \eqref{E-Motion_TensorialGeneral_Action3} for the $AdS_5$ space, the equation of motion \eqref{E-Motion_TensorialGeneral_Action1} is given by: 
\begin{equation} \label{ScalarFieldEquation}
\left[z^{3}\partial_{z}\dfrac{1}{z^{3}}\partial_{z} + \Box - \dfrac{(M_5R)^2}{z²}\right]\phi(z,\vec{x},t)=0\,,
\end{equation}
where we are disregarding the case of exotic polarizations \cite{Constable:1999gb}. 
Then, one can note that the above equation coincides with the equation of motion for a scalar field. 
Using the ansatz
\begin{equation} \label{Ansatz}
\phi(z,\vec{x},t) = e^{-iP_{\mu}x^{\mu}}\,z²\,f(z),
\end{equation}
we have the Bessel equation 
\begin{equation} \label{Bessel_Equation}
z²f''(z) + zf'(z) + [(m_{\nu,k}\,z)^{2} - \nu^2]f(z) = 0,
\end{equation}
where
\begin{equation} \label{Besselorder}
\nu² = (M_5R)^2 + 4.
\end{equation}
So, the complete solution for $ \phi(z,\vec{x},t) $ reads
\begin{equation} 
\phi(z,\vec{x},t) = C_{\nu,k}\,e^{-iP_{\mu}x^{\mu}}\,z²\,J_{\nu}({m_{\nu,k}\,z}) + D_{\nu,k}\,e^{-iP_{\mu}x^{\mu}}\,z²\,N_{\nu}({m_{\nu,k}\,z}), 
\end{equation}
where $C_{\nu,k}$ and $D_{\nu,k}$ are normalization constants, $J_\nu(\omega) $ and  $N_\nu(\omega) $ are the Bessel and Neumann functions, respectively. Since we are interested in regular solutions inside AdS space, we are going to disregard the Neumann solution. 
In this particular case, the solution  for $ \phi(z,\vec{x},t) $ becomes: 
\begin{equation} \label{ScalarFieldSolution}
\phi(z,\vec{x},t) = C_{\nu,k}\,e^{-iP_{\mu}x^{\mu}}\,z²\,J_{\nu}({m_{\nu,k}\,z}), 
\end{equation}
where $ m_{\nu,k} $ are the masses of glueball states and $ k=2,3,..., $ represents the radial excitations, with $ k=1 $ being the ground state. Since we are interested in calculating the masses of the even spin glueballs $ J^{PC} $, with $ P=C=+1 $, without radial excitations, we will omit this index $k$ from now on.

The twist of an operator is the difference between its conformal dimension $(\Delta)$ and its spin $(J)$. 
As discussed in \cite{Brower:2006ea} the pomeron is a twist two object, so that: 
\begin{equation} \label{Twist_Two condition}
\Delta - J = 2.
\end{equation}

Now, we consider a spin $ J $ operator in the four-dimensional space, with conformal dimension $\Delta$, denoted by $ \mathcal{O}_{\Delta} $ and constructed in the following way \cite{deTeramond:2005su, Ballon-Bayona:2015wra}:
\begin{equation} \label{OperatorJ}
\mathcal{O}_{\Delta} \sim \mathrm{SymTr}(F_{\beta\alpha_1}D_{\alpha_2}...D_{\alpha_{J-1}}F^{\beta}_{\alpha_J}),
\end{equation} 
where ``$ \mathrm{SymTr} $'' means symmetrized trace and $ D_{\alpha_{J}} $ are the covariant derivatives. 
This operator has twist two associated with the Pomeron 
and will be related to even spin $J$ glueball states. 

From the AdS$_5$/CFT$_4$ dictionary, one has for the five dimensional mass $M_5$ of a symmetric second-rank tensor field \cite{Polishchuk:1999nh,Naqvi:1999va,Mueck:1998ug}: 
\begin{equation} \label{AlternativeMassrelation}
(M_5R)² = \Delta(\Delta-4)  = J² - 4,
\end{equation}
where we used the twist two condition \eqref{Twist_Two condition}. One can note that this relation also holds for scalar fields \cite{Witten:1998qj, Aharony:1999ti}.

Using the above mass relationship in \eqref{Besselorder}, we obtain that the order of the Bessel function in \eqref{ScalarFieldSolution} is given by
\begin{equation}
\nu = J \,.
\end{equation}
Imposing boundary conditions on the wave functions given by eq. \eqref{ScalarFieldSolution}, we find for the Dirichlet case
\begin{equation} \label{Dirichlet_bc}
m_{\nu} = \dfrac{\lambda_{\nu}}{z_{max}}; \quad J_{\nu}(\lambda_{\nu}) = 0,
\end{equation} 
and for the Neumann case
\begin{equation} \label{Neumann_bc}
(2-\nu)J_{\nu}(\gamma_{\nu}) + \gamma_{\nu}J_{\nu-1}(\gamma_{\nu}) = 0,
\end{equation}
where
\begin{equation}
m_{\nu} = \dfrac{\gamma_{\nu}}{z_{max}}.
\end{equation}

In order to calculate the glueball masses, we have to choose $ z_{max} $. A natural choice is to associate $ z_{max} $ with the mass of some glueball state, usually the lightest state. 

In the works \cite{BoschiFilho:2005yh}, the scalar glueball state $0^{++}$ mass was used as an input, and \cite{Capossoli:2013kb}, the vector glueball state $1^{--}$ mass was used as an input. Here we are going to use the tensor glueball state $2^{++}$ mass as an input, once in this paper we are dealing with a massive symmetric second rank tensorial field.
Then, from the lattice \cite{Meyer:2004gx}, we find for Dirichlet boundary conditions 
\begin{equation}
z_{max}^{D} = 2.389 \: \mathrm{GeV}^{-1}, 
\end{equation} 
and for Neumann boundary conditions 
\begin{equation}
z_{max}^{N}= 1.782 \: \mathrm{GeV}^{-1}. 
\end{equation} 

In the next section, we will use these values of $z_{max}$ to calculate the other even spin glueball state masses with $ J^{PC} $, $ P=C=+1 $ and we will present our results for the masses for both Dirichlet and Neumann boundary conditions. Furthermore, we will obtain the Regge trajectories for the Pomeron in both cases.

\section{Pomeron Regge Trajectory and Results}

In the pre-QCD age hadron-hadron scattering processes at high energy, such as, $1 + 2 \rightarrow 3 + 4$, were treated using S-matrix approach. Due to its properties, i.e., Lorentz invariant, unitarity and analyticity (regarded as complex variables), emerged in this scenario the Regge theory to deal with those hadronic processes. Within this theory, the scattering amplitude can be written as a partial wave expansion, so that:
\begin{equation}
{\cal A}(s,t) = \sum_{\ell = 0}^\infty (2\ell +1) a_{\ell}(s) P_{\ell}\left( 1 + \frac{2s}{t - 4 m^2}\right) ,
\end{equation}
\noindent where $s$ and $t$ are the Mandelstam variables related as center-of-mass energy and obey $s + t + u = \sum_{i= 1}^4 m_i^2$. The functions $a_{\ell}(s)$ are the partial wave amplitudes.  $P_{\ell}(t,s)$ are the Legendre polynomials of degree $\ell$ and in this case, all hadrons have the same mass, so that $m_i = m$.

Within the Regge limit, that means, $s \rightarrow \infty$ with fixed $t$, ${\cal A}(s,t)$ can be read as:
\begin{equation}
{\cal A}(s, t) \sim f_{\ell}(t) s^{\ell} .
\end{equation}

Assuming that $f_{\ell}(t)$ has  a singularity structure given by
\begin{equation}
f_{\ell}(t) = \frac{\gamma(t)}{\ell - \alpha_R(t)},
\end{equation}
\noindent where $\gamma(t)$ carries all the information about incoming and outgoing particles and that it does not depends on $s$. 

These singularities or poles in the partial wave $t$-channel of the scattering process are known as Regge poles or reggeons.

The function $\alpha_R(t)$ has the following form:
\begin{equation}\label{reggeon}
\alpha_R(t) = \alpha_R(0) + \alpha'_R t,
\end{equation}
\noindent known as the reggeon or Regge trajectory with $\alpha_R(0)$ and $\alpha'_R$ constants.

Also within Regge theory one can relate the spin $J$ and mass $m$, for particles with same quantum numbers, such as $J = \alpha_R(t)$ and $m^2 = t$. Now one can rewrite the Regge trajectory as:
\begin{equation}
J(m²) = \alpha_0 + \alpha'\,m².
\end{equation}

For our purposes, we are interested in the reggeon with intercept $\alpha_0 \approx 1$, called pomeron. In the Chew-Frautschi plane, even spin glueball states lie on the pomeron Regge trajectory.

The widely accepted Regge trajectory related to the pomeron is the so-called soft Pomeron which is presented by Donnachie and Landshoff in  \cite{Donnachie:1984xq, Donnachie:1985iz} and has trajectory of the form:
\begin{equation} \label{Pomeron_trajectory}
J(m²) \approx 1.08 + 0.25\,m².
\end{equation}

After this brief discussion about the pomeron, in this section, we will present our results for the masses of the even spin glueball states for both Dirichlet and Neumann boundary conditions. From these masses we will build the Regge trajectories for the pomeron. 

In table \ref{tab1}, we show the results for the masses of glueball states $ J^{PC} $, with $ P=C=+1 $ and even $ J $ from the hardwall model using Dirichlet boundary condition and twist two operator. 
The 2$^{++}$ state mass is taken as an input from lattice \cite{Meyer:2004gx}. These masses are comparable with lattice results \cite{Chen:2005mg,Gregory:2012hu,Liu:2001wqa,Meyer:2004gx,Meyer:2004jc,Morningstar:1999rf,Lucini:2010nv} and with other approaches using the AdS/QCD models \cite{BoschiFilho:2005yh,Capossoli:2015ywa, Capossoli:2016kcr, Capossoli:2016ydo}.

\begin{table}[h]
	\begin{tabular}{|c|c|}
		\hline $J^{PC}$ & Mass (GeV) \\ 
		\hline $2^{++}$ & 2.150 \\ 
		\hline $4^{++}$ & 3.176 \\ 
		\hline $6^{++}$ & 4.159 \\ 
		\hline $8^{++}$ & 5.117 \\ 
		\hline $10^{++}$& 6.059  \\ 
		\hline 
	\end{tabular}
	\centering
	\caption{Glueball states $ J^{PC} $, $ P=C=+1 $, using Dirichlet boundary condition from eq. \eqref{Dirichlet_bc} and $z_{max}^{D} = 2.389 \: \mathrm{GeV}^{-1}$. The mass of $ 2^{++} $ is an input from the lattice \cite{Meyer:2004gx}.}
	\label{tab1}
\end{table} 

From the glueball states $2^{++}$, $4^{++}$, $6^{++}$, $8^{++}$ and $10^{++}$ in Table \ref{tab1} and using a computational linear regression method, which gives us the uncertainties related to the slope ($\alpha ' $) and intercept ($\alpha_0$) parameters, one gets the corresponding approximate linear Regge trajectory:
\begin{equation}\label{regge1}
J(m²) = (1.34 \pm 0.39)  + (0.25  \pm 0.01) m^2. 
\end{equation}

This Regge trajectory is in agreement with the one given by \eqref{Pomeron_trajectory} for the pomeron. The errors appearing in this equation come from linear regression fit.

\begin{figure}[h] 
  \centering
  \includegraphics[scale = 0.6]{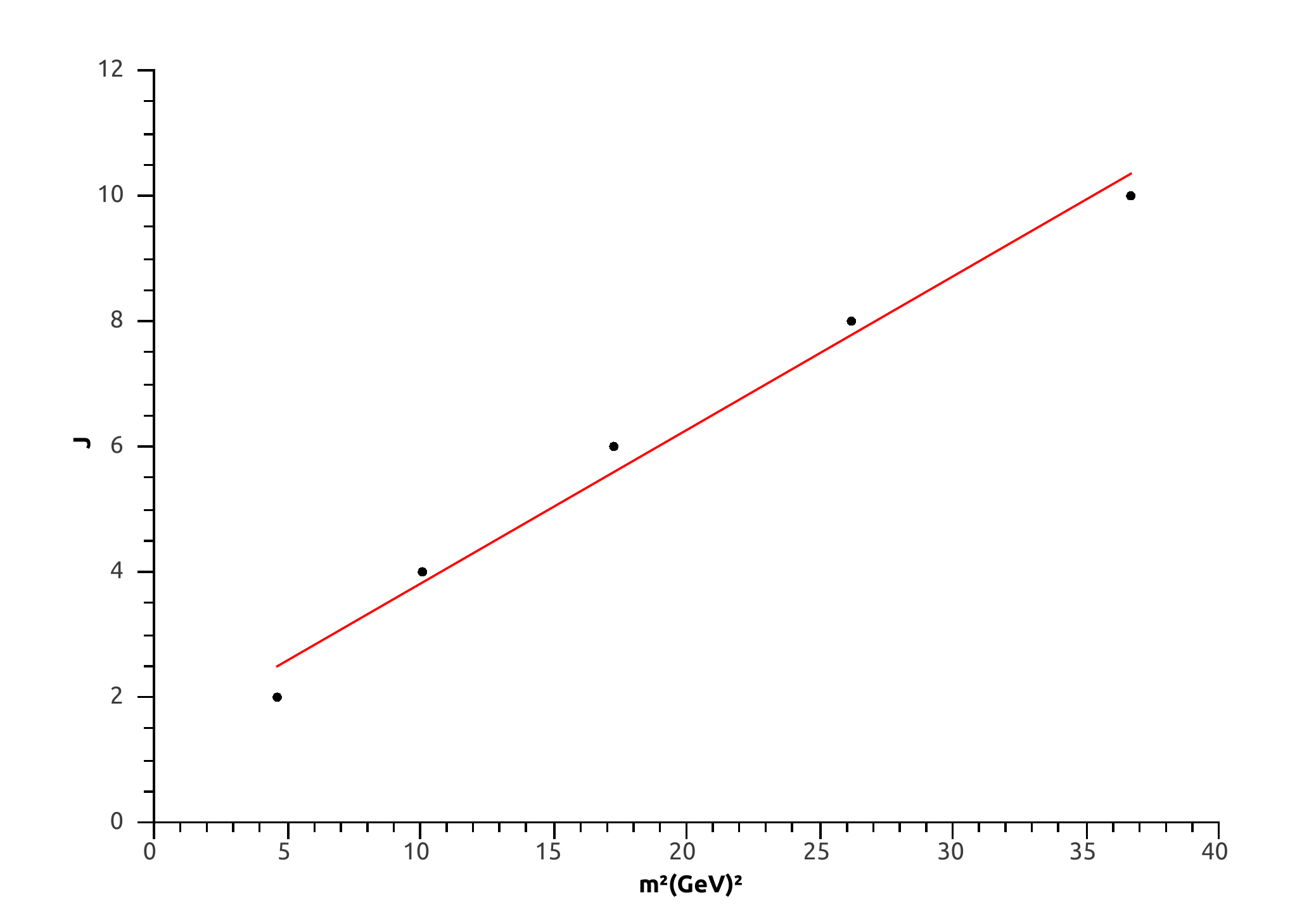} 
\caption{Glueball masses (dots) from table \ref{tab1}, using Dirichlet boundary condition. We also plot an approximate linear Regge trajectory, corresponding to Eq.(\ref{regge1}) representing the pomeron.}
\label{fig1}
\end{figure}

In figure \ref{fig1}, we present the Chew-Frautschi plot $J\times m^2$ with the masses of glueball states from table  \ref{tab1} and also show the linear Regge trajectory given by eq. \eqref{regge1} related to the pomeron. 

For another set of glueball states, in Table \ref{tab1}, for exemple, $2^{++}$, $4^{++}$ and $6^{++}$, one can find the following approximate linear Regge trajectory, given by:
\begin{equation}\label{regge1_1}
J(m²) = (0.65 \pm 0.30)  + (0.31  \pm 0.02) m^2. 
\end{equation}
This Regge trajectory is not in agreement with the one given by \eqref{Pomeron_trajectory} for the pomeron taking into account the uncertainties given by the linear regression method.

In table \ref{tab2}, we show the results for the masses of glueball states $ J^{PC} $, with $ P=C=+1 $ and even $ J $ from the hardwall model using Neumann boundary condition and twist two operator. 
The 2$^{++}$ state mass is taken as an input from lattice \cite{Meyer:2004gx}. These masses are also comparable with lattice results \cite{Chen:2005mg,Gregory:2012hu,Liu:2001wqa,Meyer:2004gx,Meyer:2004jc,Morningstar:1999rf,Lucini:2010nv} and with other approaches using the AdS/QCD models \cite{BoschiFilho:2005yh,Capossoli:2015ywa, Capossoli:2016kcr, Capossoli:2016ydo}.

\begin{table}[h]
	\begin{tabular}{|c|c|}
		\hline $J^{PC}$ & Mass (GeV) \\ 
		\hline $2^{++}$ & 2.150 \\ 
		\hline $4^{++}$ & 3.356 \\ 
		\hline $6^{++}$ & 4.546 \\
		\hline $8^{++}$ & 5.725 \\ 
		\hline $10^{++}$& 6.899 \\ 
		\hline 
	\end{tabular}
	\centering
	\caption{Glueball states $ J^{PC} $, $ P=C=+1 $, using Neumann boundary condition from eq. \eqref{Neumann_bc} and  $z_{max}^{N}= 1.782 \: \mathrm{GeV}^{-1}$. The mass of $ 2^{++} $ is an input from the lattice \cite{Meyer:2004gx}.}
	\label{tab2}
\end{table}

For the complete set of states found in Table \ref{tab2}, {\it i. e.}, $2^{++}$, $4^{++}$, $6^{++}$, $8^{++}$ and $10^{++}$, we find the approximate linear Regge trajectory:
\begin{equation}\label{regge2_1}
J(m²) = (1.74 \pm 0.44)  + (0.18  \pm 0.01) m^2. 
\end{equation}
where we used a computational linear regression method, which gives us the uncertainties related to the slope ($\alpha ' $) and intercept ($\alpha_0$) parameters. Once again, taking into account these uncertainties, this Regge trajectory has a poor agreement with the pomeron one given by \eqref{Pomeron_trajectory}.

Now, for example, using the glueball states $2^{++}$, $4^{++}$ and $6^{++}$ in Table \ref{tab2} one gets the corresponding approximate linear Regge trajectory:
\begin{equation}\label{regge2}
J(m²) = (0.99 \pm 0.34)  + (0.25  \pm 0.02) m^2. 
\end{equation}
This Regge trajectory is in good agreement with the one given by \eqref{Pomeron_trajectory} for the pomeron. The errors appearing in this equation come from linear regression fit. 

\begin{figure}[h] 
  \centering
  \includegraphics[scale = 0.6]{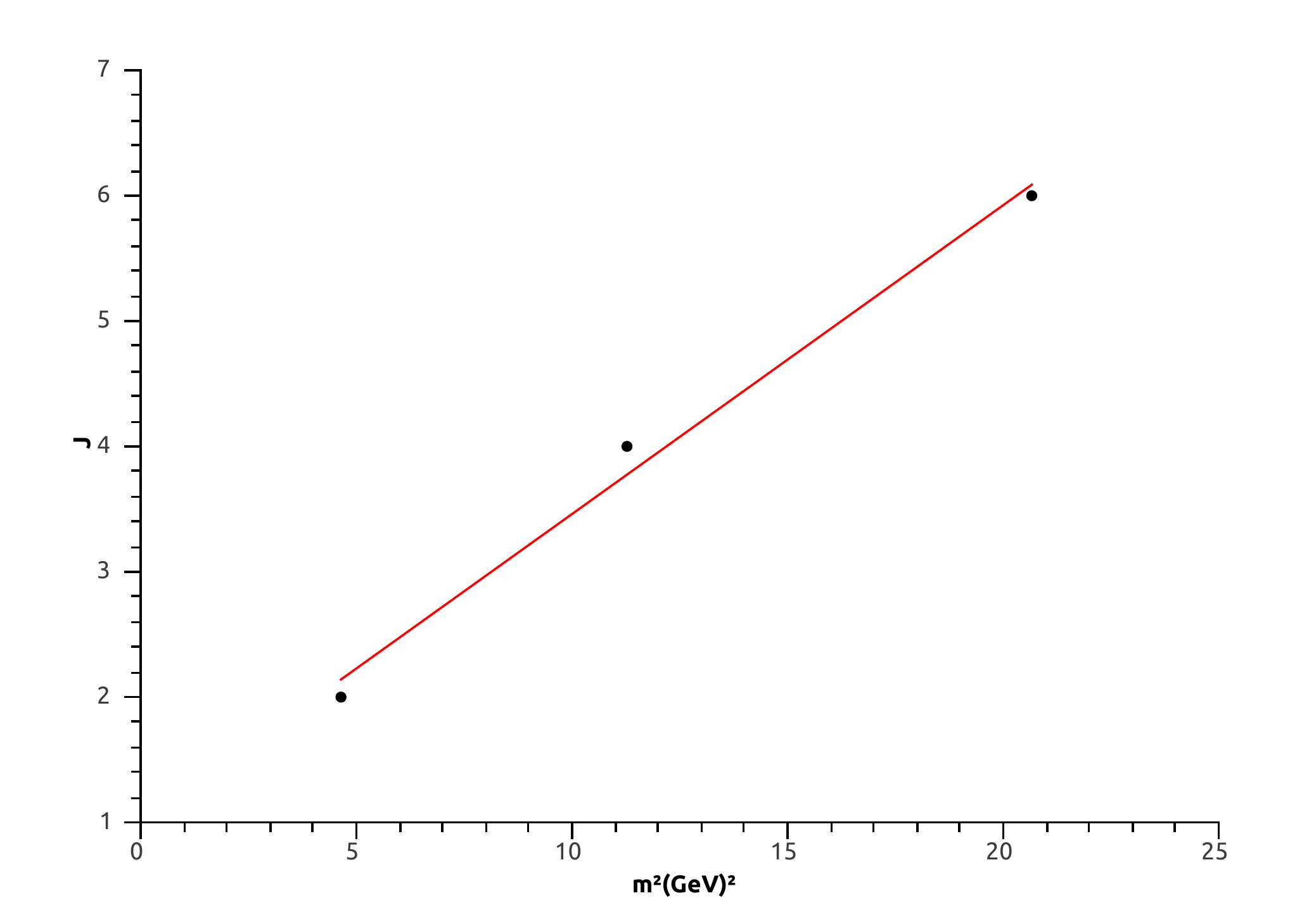} 
\caption{Glueball masses (dots) of the states  $2^{++}$, $4^{++}$, and $6^{++}$  from table \ref{tab2}, using Neumann boundary condition. We also plot an approximate linear Regge trajectory, corresponding to Eq.(\ref{regge2}) representing the pomeron.}
\label{fig2}
\end{figure}

In figure \ref{fig2}, we present the Chew-Frautschi plot $J\times m^2$ with the masses of glueball states from table  \ref{tab2} using only the states $2^{++}$, $4^{++}$, and $6^{++}$ and also show the linear Regge trajectory given by eq. \eqref{regge2} related to the pomeron. 

Note that in order to obtain the trajectory for the pomeron, we are looking for the best linear fit of the glueball states. Due to this, the number of states used in both boundary conditions are not the same.

It is worthy to say some words about the important case of the scalar glueball state $0^{++}$ and its radial excitations. 
As  the $0^{++}$ state is associated with the operator $Tr F^2$ which corresponds to a twist {\it four} operator \cite{Witten:1998qj},
we did not compute this state and its excitations in this work since we are dealing only with the twist two operator. 
 Furthermore, it is still inconclusive whether the scalar glueball state contribute, or not, to the pomeron Regge trajectory. However there are some works in holography that dealt with the scalar glueball and its radial excitations, for instance, within hardwall approach \cite{BoschiFilho:2002ta,BoschiFilho:2002vd}, and within the softwall approach \cite{Capossoli:2015ywa, BoschiFilho:2012xr, Li:2013oda, Chen:2015zhh, FolcoCapossoli:2016ejd}. In some of these works, the authors construct explicitly the Regge trajectory for the $0^{++}$ state and its radial excitations.

\section{Comments and Conclusions}

In this work we have computed the masses of even spin glueball states $ J^{PC} $, $ P=C=+1 $, 
 and obtained the Pomeron Regge trajectory using an AdS/QCD model known as the hardwall with twist two operators,  starting with massive symmetric second rank tensorial field. 
 The masses obtained are comparable with lattice results \cite{Chen:2005mg,Gregory:2012hu,Liu:2001wqa,Meyer:2004gx,Meyer:2004jc,Morningstar:1999rf,Lucini:2010nv} and also with other holographic models \cite{BoschiFilho:2005yh,Capossoli:2015ywa, Capossoli:2016kcr, Capossoli:2016ydo}.
Our results for the Regge trajectories for the pomeron using Dirichlet and Neumann boundary conditions are in agreement with experimental data \cite{Donnachie:1984xq, Donnachie:1985iz}. 

Note that in this approach with twist two operators, if one tries to include the scalar glueball state in this description
it would have conformal dimension $\Delta = 2$ which is in disagreement with AdS/CFT correspondence 
 \cite{Gubser:1998bc,Witten:1998qj,Aharony:1999ti}. Also, according to lattice studies the scalar glueball should not belong to the pomeron trajectory \cite{Meyer:2004gx,Meyer:2004jc}. 

One can also note that the outcome of the hardwall model with Neumann b.c. produce a Regge trajectory in better agreement with the literature than the Dirichlet b.c. as one can see comparing eqs.  \eqref{regge2} and \eqref{regge1} with {\eqref{Pomeron_trajectory}. This is also in agreement with the behavior which was originally found in ref. \cite{BoschiFilho:2005yh} for which the Neumann b.c. gives better results than for Dirichlet b.c. 

The comparison of the Regge trajectories obtained here for the Neumann b.c. and Dirichlet b.c.  with the results from the softwall model \cite{Capossoli:2016kcr, Capossoli:2016ydo} also seems to favor the Neumann b.c. 

Another important difference between our results for Neumann and Dirichlet b.c. lies on the fact that the Regge trajectory built with Neumann b.c. includes only the states $2^{++}$, $4^{++}$, and $6^{++}$ while the one for Dirichlet b.c. includes also the states $8^{++}$ and $10^{++}$. Then, we conclude that the Neumann b.c. gives a simpler formulation of the Regge trajectory than the one with Dirichlet b.c., since it was not possible to obtain a reasonable Regge trajectory using Dirichlet b.c. with just the three lighter states. 


\begin{acknowledgements}
D. M. R. thanks CNPq (Brazilian agency) for financial support. E. F. C. thanks CPII-PROPGPEC for partial financial support. H.B.-F. also thanks CNPq for partial financial support. 
\end{acknowledgements}

\end{document}